\documentclass{mn2e}
\usepackage{epsfig}
\usepackage{graphicx}

\title[Mass accretion in pre-main sequence stars]
{The $\dot{M}-M$ relationship in pre-main sequence stars}

\author[ C. J. Clarke \& J. E. Pringle]
{C. J. Clarke$^1$  \& J. E. Pringle$^1$\\
$^1$Institute of Astronomy, Madingley Road, Cambridge, CB3 0HA, UK\\
}

\begin{document}

\maketitle

\begin{abstract}

We examine the recent data and analysis of Natta et al. concerning the
accretion rate on to young stars as a function of stellar mass, and
conclude that the apparently steep dependence of accretion rate on
mass is strongly driven by selection/detection thresholds. We argue
that a convincing demonstration of a physical relationship between
accretion and stellar mass requires further studies which, as is the
case for Natta et al., include information on upper limits, and which
quantify the possible incompleteness of the sample, at both low and
high accretion rates. We point out that the distribution of detections
in the ($M , \dot{M}$)-plane can in principle be used to test
conventional accretion
disc evolutionary
models, and that higher sensitivity observations might be able to
test the hypothesis of accelerated disc clearing at late times.

\end{abstract}


\section{Introduction}

There appears to be increasing acceptance that the accretion rate,
$\dot{M}$, onto pre-main sequence stars, although it shows a large
scatter, correlates roughly as the square of the stellar mass, $M$
(Muzerolle et al 2003; Natta et al 2004; Calvet et al 2004; Muzerolle
et al 2005; Mohanty et al 2005; Natta et al 2006). From a theoretical
point of view this is a somewhat surprising finding in that it appears
to indicate that the accretion processes at different masses do not
scale simply with mass.  It has been interpreted variously as
indicating that accretion is Bondi-Hoyle accretion in a uniform
environment (Padoan et al., 2005) or as giving us information about
the initial conditions established when the discs form (Alexander \&
Armitage, 2006; Dullemond, Natta and Testi, 2006).

The recent compilation of Natta et al. (2006) however permits us to
examine this claim more quantitatively than has been possible
hitherto. Our analysis here suggests that the distribution of stars in
the plane of accretion rate versus mass is largely bounded by
detection and selection thresholds and that the claimed steep
relationship between accretion rate and mass is then an inevitable
consequence of these thresholds. We, however, stress that we in no way
rule out the possibility that mean accretion rate is a steep function
of stellar mass, but instead set out the conditions that must be
satisfied by future datasets before this claim can be taken at face
value.

\section{Measurements, upper limits and correlations}

The possibility that there is a strong correlation between accretion
rate and stellar mass was noted by Muzerolle et al. (2003), who
assembled accretion rate estimates obtained through a variety of
diagnostics (fitting of emission line profiles, veiling measurements,
determination of U band excesses: see Figure 8 of Muzerolle et al.,
2003). In order to make a critical assessment of this claim, however,
it is necessary to work with a well--defined. homogeneous sample for
which the accretion rate is measured in a uniform manner. The recent
dataset of Natta et al. (2006) in the core of $\rho$ Oph satisfies
these criteria and in addition, and very valuably, lists the upper
limits for non-detections.

\begin{figure}

\includegraphics[width=0.5\textwidth,angle=0]{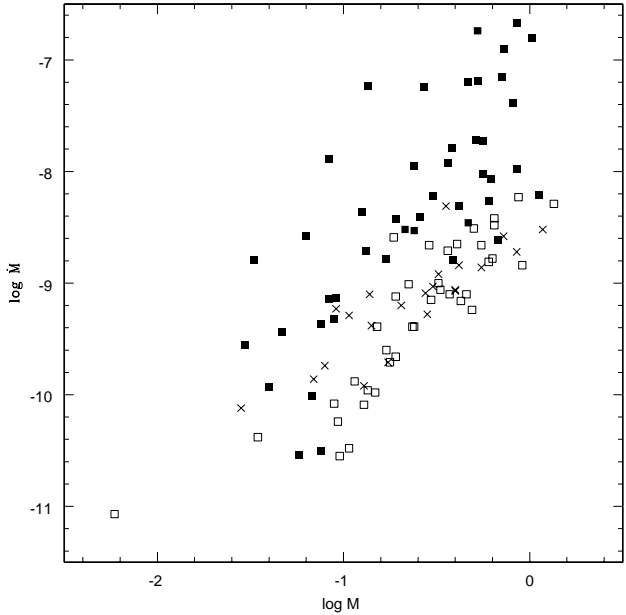}

\caption{Accretion rate is plotted against stellar mass for data
points derived from Pa$\beta$ in Natta et al. (2006). Filled squares
are measurements, and open squares are upper limits for Class II
objects. Crosses are upper limits for Class III objects.}

\label{mdotm}

\end{figure}

In Figure~\ref{mdotm}, we plot the detected measurements of accretion
rate, together with the upper limits, as a function of stellar mass
(see also Natta et al., 2006, Figure 2).~\footnote {Note that in the
interests of uniformity of analysis we use only measurements based on
Pa$\beta$ and omit the eleven objects for which only Br$\gamma$
measurements are available}.  It is immediately evident, first, that
the upper limits also correlate with mass in the same way as the
detections, and, second, that the upper limits and detections overlap
at each value of $M$. It is therefore clear that the observed
$\dot{M}-M$ relation is being strongly driven by detectability limits
in the $(\dot{M} , M)$-plane.

\begin{figure}


\includegraphics[width=0.5\textwidth,angle=0]{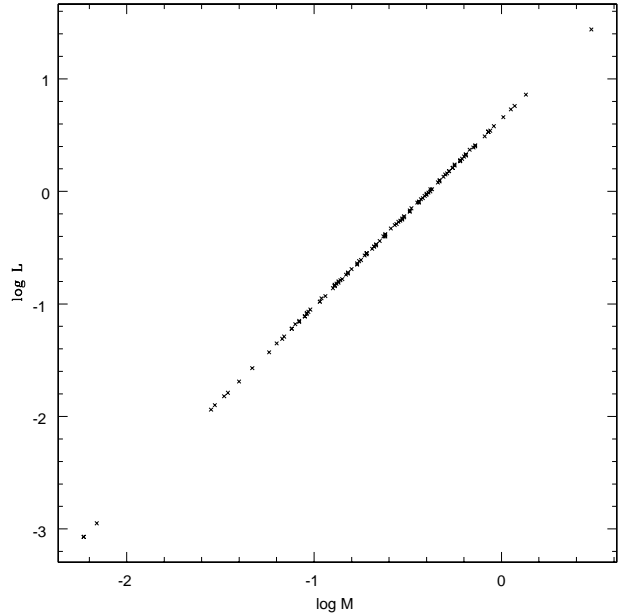}

\caption{The assumed luminosity-mass relation for the stars in the
Natta et al (2006) sample.}

\label{LM}

\end{figure}

\begin{figure}


\includegraphics[width=0.5\textwidth,angle=0]{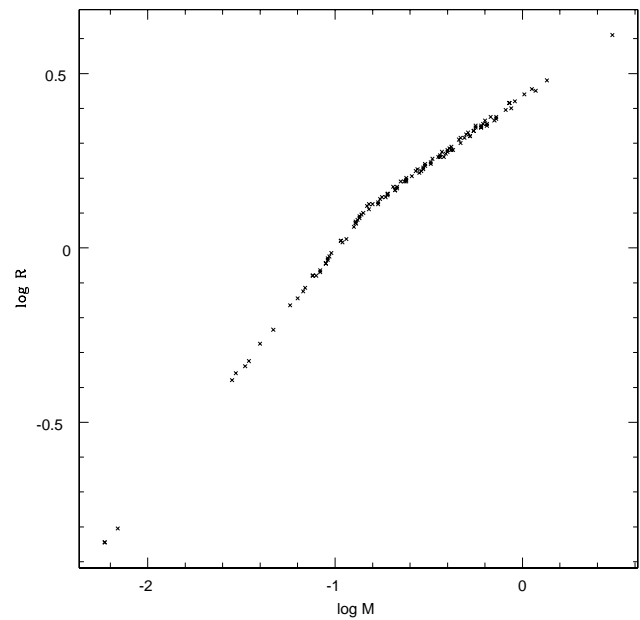}

\caption{The assumed radius-mass relation for the stars in the Natta et
al. (2006) sample.}

\label{RM}

\end{figure}

The reason for the strong correlation between detectability of
accretion rate and mass is easy to understand and results from a
combination of two effects. First, Natta et al. (2006) make the
reasonable assumption the the observed stars fall on an
isochrone. This assumption implies that there are therefore strong
correlations between luminosity, $L$, and mass (Figure~\ref{LM}) and
between radius, $R$, and mass (Figure~\ref{RM}) for the stars in the
dataset. We should note that even if this assumption is relaxed, and
an attempt is made to estimate masses and radii by placing the stars on
pre-main sequence tracks in the HR diagram, because all the stars
observed are of a similar age, all other datasets show similar,
although not so tight, correlations. For the Natta et al. (2006)
dataset we find roughly that $L \propto M^{1.7}$ and $R \propto
M^{0.6}$. Second, Natta et al. (2006) use the equivalent width of the
Pa$\beta$ line as their primary indication of accretion rate. What
this means in effect is that, in line with all other methods,
accretion rate is estimated by measuring extra flux of some kind over
and above that expected for the stellar photospheric (including
chromospheric) flux for the star in question. Thus to first order, all
estimates of accretion rate rely essentially on estimating the
dimensionless quantity

\begin{equation}
k = \frac{L_{\rm acc}}{L},
\end{equation}
where
\begin{equation}
L_{\rm acc} = \frac{G M \dot{M}}{R},
\end{equation}
is the accretion luminosity.

>From these relations we see that

\begin{equation}
\dot{M} = \frac{k R L}{GM}.
\end{equation}

\begin{figure}


\includegraphics[width=0.5\textwidth,angle=0]{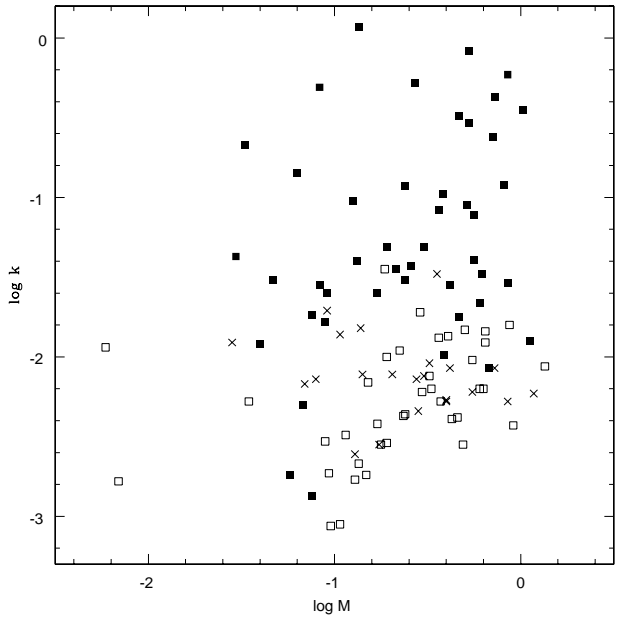}

\caption{The value of $k = L_{\rm acc}/L$ plotted against stellar mass
for data points derived from Pa$\beta$ in Natta et al. (2006). Filled
squares are measurements, and open squares are upper limits for Class
II objects. Crosses are upper limits for Class III objects.}

\label{kM}

\end{figure}

We plot $k$ as a function of $M$ for the Natta et al. (2006) data in
Figure~\ref{kM} (see also their Figure 2). The plot displays
considerable scatter in $k$ at a given mass (more than two orders of
magnitude). The regions of the plot that are filled with points are
simply bounded by detection limits - e.g. the observed values of $k$
are all less than unity since accretion rates are not readily
determinable in the case that that the accretion rate exceeds the
stellar luminosity (Hartmann et al., 1998). In addition, there is an
apparent mild slope to the lower boundary of the distribution, i.e. an
absence of detections in the lower right portion of the diagram. This
however is also explicable as a sensitivity issue, since the upper
limits follow the same trend.  We therefore conclude that the
detectable region of $(k,M)$ parameter space is rather uniformly
populated with detections.  At fixed $k$, the correlations between
$M$, $R$ and $L$ imply $\dot{M} \propto M^{1.3}$.  We can therefore
see that if the detectable region of $(M,k)$ parameter space were
roughly uniformly populated with detections, in some range $k_1 < k <
k_2$ and $M_1 < M < M_2$, then the tightness of the $R-M$ and $L-M$
correlations would give a spurious correlation between $\dot{M}$ and
$M$.~\footnote{This is analagous to the well known phenomenon that
even if the two quantities $A$ and $B$ are uncorrelated, the
quantities $A f(B)$ and $B$ can in fact appear correlated.}  In our
case because there is a mild positive slope to the lower bound of the
detectable region in the ($M,k$)-plane, the detected systems
necessarily exhibit a correlation between accretion rate and mass that
is somewhat steeper than $\dot{M} \propto M^{1.3}$.

\section{Discussion}

\subsection{The lack of high accretion rates at low masses}

We have argued above that that maximum values of accretion rate at all
masses all correspond roughly to the case that the accretion
luminosity is comparable with the stellar luminosity $(k \approx 1)$,
and that this could just be an observational selection effect. For
example, in the case of Taurus, Kenyon and Hartmann (1995) classified
six stars as `continuum stars', where the level of veiling was so high
as to fill in the photospheric absorption lines. It is an important
matter to establish the number of systems that could have been missed
in the present sample.

If, in reality, there are few objects that have been systematically
excluded on account of high accretion rate ($k > 1$), then there
appears to be lack of systems in the upper left of
Figure~\ref{mdotm}. In this case, we need to enquire whether this lack
is statistically significant given the present sample. In other words,
we need to determine if the null hypothesis (viz. that the
distribution of $\dot{M}$ is independent of $M$) is contradicted by
the data.

We analyse this issue by splitting the data mass into two mass bins,
with $M < M_{\rm c}$ and $M > M_{\rm c}$, for some mass $M_{\rm c}$.
and then using a modified version of the Kolmogorov--Smirnov (KS)
test. If we had accretion rate measurements for all the objects
studied (rather than just upper limits for some), then we would apply
the KS test by computing the cumulative distributions of $\dot{M}$ in
the two mass bins, by finding the largest difference between these two
distributions, and by then using the KS statistic to assess the
significance of this difference. In the current case, however, a
number of the data points are upper limits, rather than actual
measurements. But since the the upper limits in Figure 1 are all at
accretion rates $\log \dot{M} < -8.1$, we can still measure the
greatest difference between the two cumulative distributions in the
domain $\log \dot{M} > -8.1$, and then use the KS test to assess the
significance of that difference.  Note that since the actual
difference (had we been able to measure all the values of $\dot{M}$)
is greater than this, this procedure in fact underestimates the
significance of the difference between the two distributions. We have
undertaken this analysis using a number of trial values of $M_{\rm c}$
and find that if we split the sample at $\log M_{\rm c} = - 0.5$
(i.e. at $M_{\rm c}= 0.3 M_\odot$) the KS probability is around
$0.04$, i.e. we can rule out the null hypothesis at about the $2
\sigma$ level. Small changes in the value of $M_{\rm c}$ (i.e. by
$0.1$ dex) however increases the KS probability and renders the
difference insignificant at the $2 \sigma$ level.

We therefore conclude that, {\it if} there are in reality no systems
that have been systematically excluded on account of excessively large
accretion rates, then the difference in accretion rate distribution at
high and low mass is marginally statistically significant.~\footnote
{If, however, we perform a similar analysis on the quantity $\dot
M/M$, implying a null hypothesis that $\dot{M} \propto M$, we find no
significant difference between the distributions in different mass
bins. Thus $\dot{M} \propto M$ is consistent with the data.} This
result is independent of any issues relating to the mass dependence of
the lower detection limits, owing to the way that we have included the
upper limits in our analysis.  Larger samples should be able to
improve the statistical significance of this result and make it less
sensitive to exactly where the mass cuts are taken.  We stress that
such analysis can only be conducted using samples which, like that of
Natta et al. (2006), both contain data acquired in a uniform fashion
and which include all the information on upper limits for non-detections.

\subsection{ What does the distribution in the $\dot{M} - M$ plane tell us?}

The distribution of detections in the $\dot{M} - M$ plane yields
information about the relative amounts of time that systems spend in
various regions of that plane. In fact, since the mass of a star
changes negligibly during the Class II phase, this plot is mainly a
diagnostic tool for the evolution of $\dot{M}$ during this
phase. Simple disc models generally predict that $\dot{M}$ should
decline as a power law of the time, in which case the  number of
objects found in equal intervals of $\log
\dot{M}$ should increase somewhat at lower
$\dot {M}$..~\footnote{There are however individual cases where the
accretion timescale ($t_{\rm acc} = M_{\rm disc}/\dot{M}$) is
apparently less than the stellar age, a result that is most easily
explained in terms of fluctuating accretion rates (see Scholz and
Jayawardhana, 2006, and Littlefair et al., 2004, for evidence that
some T Tauri stars show pronounced changes in spectral
characteristics).}  In principle, one could split the sample according to
stellar mass and then use the distribution of accretion rates
to test the hypothesis of power law decline of
$\dot {M}$ with time, and measure its exponent. The
current dataset is however too sparse to permit this
exercise.

The scarcity of T Tauri stars with infrared colours that are
intermediate between those of Class II and Class III sources (see
Kenyon and Hartmann, 1995; Simon and Prato, 1995) however indicates
that there may at some point be a rapid evolution in $\dot{M}$,
i.e. `two timescale' behaviour that is incompatible with a simple
power law decline (Armitage, Clarke \& Tout, 1999; Clarke, Gendrin \&
Sotomayor, 2001). This behaviour would be manifest as a lack of points
below some (possibly mass dependent) accretion rate.  There is no
evidence for such behaviour in the current dataset since the lowest
detection and upper limits are inter-mingled, and the same is true of
accretion rates based on veiling and U band measurements (see White
and Ghez, 2001). Even in the case of the most sensitive accretion rate
measurements (obtained via modeling H$\alpha$ emission lines profiles
in brown dwarfs), the lowest detected rates are within a factor two of
the estimated sensitivity limit (Muzerolle et al., 2003), which, given
the uncertainties of $\sim 3-5$ in the derived rates, cannot be taken
as evidence for a gap.

The predicted accretion rates at which rapid evolution should set in
are however very low.  In `ultraviolet switch' models for disc
dispersal (Clarke et al., 2001; Alexander, Clarke \& Pringle,
2006a,b), the accelerated dispersal occurs when the accretion rate
becomes comparable with the rate of photoevaporation by the star's
ionising flux, which is estimated to be $\sim 10^{-10} M_\odot$
yr$^{-1}$ for solar mass stars (Hollenbach et al., 1994; Alexander,
Clarke \& Pringle, 2005).  This rate scales with the square root of
the product of stellar mass and ionising flux, so that {\it if} the
ionising flux simply scales with stellar luminosity, the critical
$\dot{M}$ scales as $\propto M^{1.35}$. However the sharp drop in
chromospheric activity in late type stars (Dobler, Stix \&
Brandenburg, 2006) suggests that the critical $\dot{M}$ might scale
more strongly with mass. Evidently these accretion rates are not
measurable with current data. Alternatively, it is often assumed that
the rapid transition in infrared colours is instead associated with
grain growth in the inner disc (e.g. Sicilia-Aguilar et al., 2006), in
which case one might not expect any deviation from power law decline
in accretion rate with time at low $\dot{M}$.  Future high sensitivity
observations could thus in principle distinguish between these
possibilities. It should however be noted that, in future experiments,
the outcome should not be prejudiced by throwing out of the sample
those objects that are deemed in advance to be non-accretors (see
Muzerolle et al., 2003, and Mohanty et al., 2005, for details of
pre-selection of possible accretors based on H$\alpha$ equivalent
widths).

\section{Conclusion}

We have shown that the $\dot{M} - M$ measurements of Natta et al.
(2006) are bounded by the conditions $L_{\rm acc} \sim L$ at high
$\dot{M}$ and by a lower bound that is defined by the locus of upper
limits in this plane. The region bounded by these detection/selection
thresholds is filled rather uniformly by detections and the slopes of
these thresholds are such that the detections exhibit a steep
dependence of mean $\dot{M}$ on $M$. A conservative interpretation is
therefore that current data cannot be used to demonstrate a {\it
physical} correlation between accretion rate and stellar mass.

A number of interesting points should be borne in mind for future
investigations.

i) If the absence of systems with $L_{\rm acc} > L$ is real (i.e.
not a selection bias) then the null hypothesis that accretion rate is
independent of mass can be marginally rejected at about the $2 \sigma$
level. Larger samples are required in order to demonstrate the
statistical significance of this result.

ii) Within the regions of detectability, stars are roughly uniformly
distributed in $\log \dot{M}$. Such behaviour is compatible with a
power law decline in accretion rate with time, as predicted by
conventional accretion disc models.

iii) Higher sensitivity measurements across a range of masses could in
principle establish whether there are empty regions at low accretion
rates, corresponding to a phase of rapid disc clearing (as, for
example, in photoevaporation models).

\section{Acknowledgments} 

We thank Richard Alexander for valuable and illuminating
discussions and Antonella Natta for comments on an earlier draft
of the paper. JEP thanks IUCAA, Pune, and RRI, Bangalore, for
hospitality while this work was being initiated.

\end{document}